\documentclass{jlab}

\def\Resultphirhozeroformal{5.7\pm 0.5(\mathrm{stat}) \pm 0.8(\mathrm{syst}) ~\mathrm{fb}}
\def\Resultrhozerorhozeroformal{20.7 \pm 0.7 (\mathrm{stat})\pm 2.7(\mathrm{syst}) ~\mathrm{fb}}
\def\Resultphieta{2.1\pm 0.4 (\mathrm{stat})\pm 0.1(\mathrm{syst}) ~\mathrm{fb}}
\def\Resultphietascale{2.9\pm 0.5 (\mathrm{stat})\pm 0.1(\mathrm{syst}) ~\mathrm{fb}}
\def\Myamplitudes{|\mathrm{F}_{00}|^2:|\mathrm{F}_{10}|^2:|\mathrm{F}_{11}|^2=0.51\pm0.14(\mathrm{stat})\pm0.02(\mathrm{syst}):0.10\pm0.04 (\mathrm{stat})\pm0.01(\mathrm{syst}):0.04\pm0.03(\mathrm{stat})\pm0.00(\mathrm{syst}) }
\def\Resultrhoprhom{8.5\pm 0.7 (\mathrm{stat})\pm 1.5(\mathrm{syst}) ~\mathrm{fb}}
\def\ResultrhoprhomExtend{20.0\pm 1.6 (\mathrm{stat})\pm 3.6(\mathrm{syst}) \pm 1.7(\mathrm{ampl}) ~\mathrm{fb}}

\usepackage{relsize}
\def\babar{\mbox{\slshape B\kern-0.1em{\smaller A}\kern-0.1em
    B\kern-0.1em{\smaller A\kern-0.2em R}}}

\begin{document}
SLAC-PUB-12820 

\title{$e^+e^-$ Annihilations into Quasi-two-body Final States \\ at 10.58 GeV}

\author{Kai Yi}

\address{Stanford Linear Accelerator Center\\
Menlo Park, California 94025, USA\\
$^*$E-mail: yik@slac.stanford.edu}

\begin{abstract}
We report the first observation of $e^+e^-$ annihilations into hadronic 
states of positive $C$-parity, $\rho^0 \rho^0$  and $\phi\rho^0$. 
The angular distributions support  two-virtual-photon 
annihilation production. We also report the observations of 
$e^+e^-\rightarrow \phi\eta$ and a preliminary result on  
$e^+e^-\rightarrow \rho^+\rho^-$.

\end{abstract}

\keywords{Exclusive; High momentum; C parity; Helicity amplitude.}

\bodymatter

\section{Introduction}\label{aba:sec1}

The large datasets collected by the $B$ factories provide unique opportunities
for studying  rare processes and discovering new states. 
We report several observations of $e^+e^-$ annihilations into quasi-two-body hadronic final states with 
$C=\pm 1$ at \babar~\cite{eetovv,eetophieta,eetorhoprhom}. A new avenue for the study of hadron 
production mechanisms is opened with these observations, and a testing ground for 
QCD at the amplitude level is provided.

\section{$e^+e^-\rightarrow\rho^0 \rho^0,\phi\rho^0$}

\begin{figure}
\begin{minipage}{1.0 \textwidth}
\begin{center}
\includegraphics[width=5.5cm]{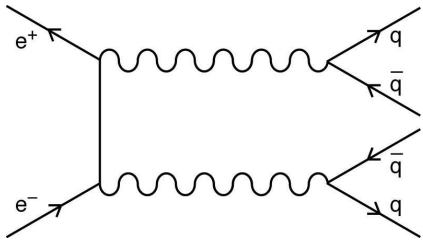}
\caption{\small The two-virtual-photon annihilation diagram. }
\label{fig:twovirtualphoton}
\end{center}
\end{minipage}
\end{figure}

The process $e^+e^-\rightarrow \mathrm{hadrons}$ at center-of-mass (c.m.) energy $\sqrt{s}$  
far below the $Z^0$ mass is dominated by annihilation via a single virtual photon, thus yielding 
final state charge-conjugation parity $C=-1$. 
The Two-Virtual-Photon-Annihilation (TVPA) process, 
depicted in Fig.~\ref{fig:twovirtualphoton},  
with positive final state C parity, has 
been ignored in incorporating  
the total hadronic cross section in $e^+e^-$ annihilations into 
calculations~\cite{g-2} of muon $g\! -$2,  and the running of the QED coupling constant, $\alpha$. 

The present analysis uses a 205 fb$^{-1}$ data sample  collected  
at the $\Upsilon(4S)$ resonance, and 20 
fb$^{-1}$ collected at c.m. energy 40 MeV lower, using the   
\babar~ detector at the SLAC PEP-II asymmetric-energy $e^+e^-$ collider. 
The \babar~ detector is described in detail elsewhere~\cite{babardetector}.

Events with four well-reconstructed charged tracks    
and net charge zero are selected. The $\chi^2$
probability of the fitted four track vertex is required to exceed 0.1\%, 
and two oppositely charged tracks  must   be identified
as pions; the other pair must  be identified as two pions or two kaons.
We accept events with four-particle invariant mass  within 170 MeV/c$^2$ of the
nominal c.m. energy.
Loose signal regions are defined by the mass ranges  
$0.5<m_{\pi^+\pi^-}<1.1$ GeV/c$^2$ and $1.008<m_{K^+K^-}<1.035$ GeV/c$^2$. 
The extracted $\rho^0\rho^0$ and $\phi\rho^0$ yields in these intervals  are 
$1243\pm 43$ and $147\pm 13$ events, 
respectively.

\begin{figure}
\begin{minipage}{1.0 \textwidth}
\begin{center}
\includegraphics[width=8.5cm]{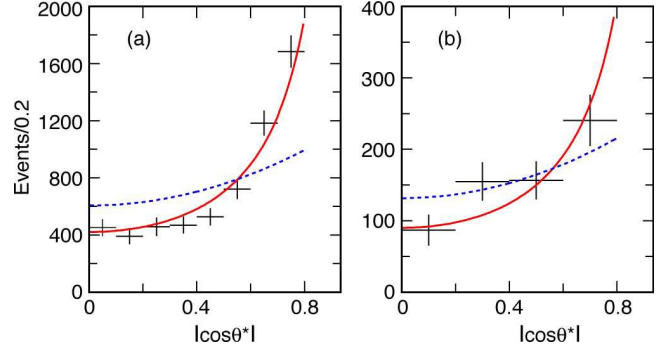}
\caption{\small 
Production angle distributions, after correction for efficiency, for a) $\rho^0\rho^0$ and b)
$\phi\rho^0$. The solid and dashed lines are the normalized 
$\frac{1+\cos^2\theta^*}{1-\cos^2\theta^*}$ and  
$1+\cos^2\theta^*$ distributions, respectively. }
\label{fig:myproduction}
\end{center}
\end{minipage}
\end{figure}

The efficiency-corrected production angular distributions are shown  in 
Fig.~\ref{fig:myproduction}, where  $\theta^*$ is  
defined as the angle between  the $\rho^0_f$ ($\phi$)  direction and 
the $e^-$ beam direction in the  c.m. frame.
The observed sharply  peaking $|\cos\theta^*|$ distributions are consistent with the 
TVPA expectation~\cite{davier}, which is approximated by: 
\begin{equation}
\label{prodangle}
\frac{d\sigma}{d\cos\theta^*}\propto\frac{1+\cos^2\theta^*}{1-\cos^2\theta^*}.
\end{equation}

For the signal mass regions defined above, and $|\cos\theta^*|<$0.8, 
we obtain the following results for the TVPA cross sections  near $\sqrt{s}=10.58$ GeV: 
\begin{eqnarray*} 
 \sigma_{\mathrm{fid}}(e^+e^-\rightarrow\rho^0 \rho^0) & = & \Resultrhozerorhozeroformal \\
 \sigma_{\mathrm{fid}}(e^+e^-\rightarrow\phi \rho^0) & = & \Resultphirhozeroformal .
\end{eqnarray*}
The measured cross sections are in good agreement with  the calculations~\cite{davier,bodwin}.
The Standard Model calculations of the anomalous magnetic moment of the muon  and 
of the QED coupling  constant rely on 
measurements of low-energy $e^+e^-$ hadronic cross sections, 
which are assumed to be entirely due to single-photon exchange. 
We have estimated the effect  due to the TVPA processes~\cite{davier} and find it to be 
small compared with the current precision~\cite{g-2}.


\section{$e^+e^-\rightarrow\phi\eta$}

The process $e^+e^-\rightarrow J/\psi \eta_c$ and other double charmonium processes are 
observed~\cite{doublecc} at rates approximately ten times larger than  expected  
from QCD-based models~\cite{nrqcd}. 
Various theoretical efforts to understand this have been made recently
~\cite{aclightcone}. 
An alternate  avenue of investigation is provided by the   process $e^+e^-\rightarrow\phi\eta$, 
which also involves 
a $\mathrm{vector-pseudoscalar} ~(\mathrm{VP})$ final state.
Different models predict different $s$  dependences for the cross section, and so it 
is interesting to investigate this  
by comparing a measurement  at $\sqrt{s}=10.58$ GeV  
to the  CLEO  measurement 
at  $\sqrt{s}=3.67$ GeV~\cite{cleophieta}.

This analysis uses  204 fb$^{-1}$  of $e^+e^-$ colliding beam data collected  
on the $\Upsilon(4S)$ resonance at $\sqrt{s}=10.58$ GeV and 20 fb$^{-1}$ collected 40 MeV below. 
Events with exactly two well-reconstructed, 
oppositely charged kaon tracks  and at least  
two well-identified photons  are selected. We fit the two tracks to a common vertex, and require the $\chi^2$
probability to exceed 0.1\%. Each photon candidate  
is  required to have a minimum laboratory energy of 500 MeV. 
Events with a reconstructed  $K^+K^-\gamma\gamma$ invariant mass    within 230 MeV/c$^2$ of the
$e^+e^-$ c.m. energy are accepted for further study.

We define the $\phi$ mass window  as  $1.008~<~m_{KK}~<~1.035$ GeV/c$^2$,  
and extract $24\pm 5$ $\phi\eta$ signal events   in the   $\phi$  mass window, with $\eta\rightarrow\gamma\gamma$.
The significance is estimated to be 6.5 sigma.

\begin{figure}
\begin{minipage}{1.0 \textwidth}
\begin{center}
\includegraphics[width=6.cm]{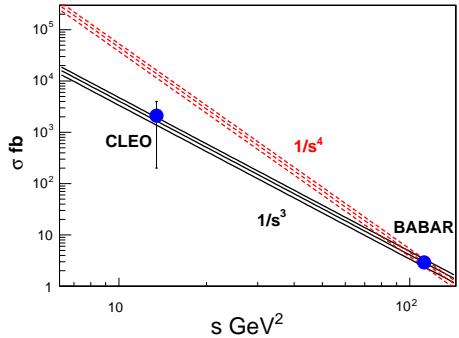}
\caption{ \small 
Cross section extrapolations 
based on \babar's measurement at $\sqrt{s}=10.58$ GeV assuming a $1/s^3$ (black) or $1/s^4$ (red) 
energy dependence. The bands show one standard deviation uncertainties in the extrapolations.
The CLEO measurement at $\sqrt{s}=3.67$ GeV is also shown.
}
\label{fig:sdependence}
\end{center}
\end{minipage}
\end{figure}

The final radiation-corrected cross section for  
$1.008~<~m_{\phi}~<~1.035 $ GeV/c$^2$   
within $|\cos\theta^*|<$ 0.8 near $\sqrt{s}=10.58$ GeV is: 
\begin{eqnarray*}
 \sigma_{\mathrm{fid}}(e^+e^-\rightarrow\phi \eta) & = & \Resultphieta .
\end{eqnarray*}
The cross section, extended to  the full range of $\cos\theta^*$ 
by assuming a $1+\cos^2\theta^*$ distribution, is: 
\begin{eqnarray*}
 \sigma(e^+e^-\rightarrow\phi \eta) & = & \Resultphietascale .
\end{eqnarray*}
There is currently no direct prediction for the cross section of this
process at this energy,
but the $e^+e^-\rightarrow\mathrm{VP}$  cross section is expected to have a $1/s^2$~\cite{gerard}
or $1/s^4$~\cite{stanrule,Chernyak} dependence in QCD-based models.
A comparison between our result and that of CLEO,
($\sigma=2.1^{+1.9}_{-1.2}\pm0.2 ~\mathrm{pb}$) at $\sqrt{s}=3.67$ GeV (continuum)~\cite{cleophieta},
favors a $1/s^3$ dependence (Fig.~\ref{fig:sdependence}).

\begin{figure}
\begin{minipage}{1.0 \textwidth}
\parbox[t]{1.0\textwidth}{\it \babar~ Preliminary  }
\begin{center}
\includegraphics[width=9cm]{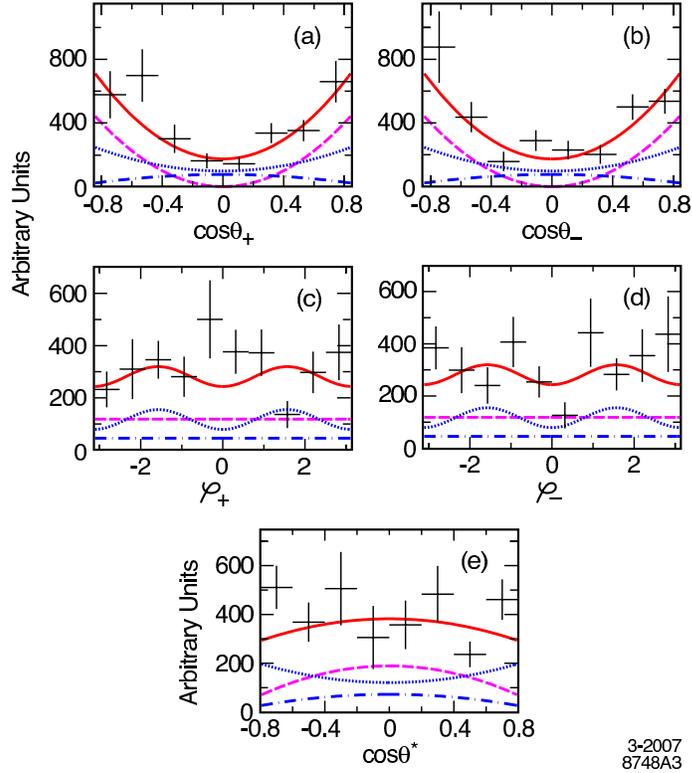}
\caption{
The s-weighted and efficiency corrected  a) $\cos\theta _+$ b) $\cos\theta _-$  
c) $\varphi_+$  d) $\varphi_-$  e) $\cos\theta^*$  
distributions for $e^+e^-\rightarrow\rho^+\rho^-$.  The magenta dashed curves show  
the contributions from $\mathrm{F}_{00}$, the 
blue dotted curves are  
$\mathrm{F}_{10}$, the blue dashed-dotted curves are $\mathrm{F}_{11}$, 
and the solid red  curves show the total result.
 }
\label{fig:angles}
\end{center}
\end{minipage}
\end{figure}

\section{$e^+e^-\rightarrow\rho^+\rho^-$ (preliminary result)}

Since charged $\rho$'s are involved, 
the $e^+e^-\rightarrow\rho^+\rho^-$ process is unlikely to occur  through TVPA~\cite{eetovv,davier,bodwin}, 
unless there is significant final quark recombination between the products of the two virtual photons, or 
unless there is  significant 
final state interaction ($e^+e^-\rightarrow\rho^0\rho^0\rightarrow\rho^+\rho^-$)~\cite{reggeo}.   
Assuming a one-photon production mechanism, 
this VV ($\rho^+\rho^-$) final state is described by three helicity amplitudes.  
A study of this reaction can then provide an experimental 
test of QCD at the amplitude 
level~\cite{stanrule}~\cite{Chernyak} through investigation of  the final states angular correlations.

This analysis uses  343 fb$^{-1}$  of $e^+e^-$ colliding beam data collected  
on the $\Upsilon(4S)$ resonance at $\sqrt{s}=10.58$ GeV and 36 fb$^{-1}$ collected 40 MeV lower.   
Events with exactly two well-reconstructed, oppositely charged tracks identified as pions and at least
two well-reconstructed $\pi^0$s  are selected. 
We fit the charged tracks to a common vertex, and require the $\chi^2$
probability to exceed 0.1\%. 
Each $\pi^0$ is reconstructed through its $\gamma\gamma$ decay channel by requiring 
the two photon invariant mass to be within the range  
[0.1,0.16] GeV/c$^2$, and then constraining its mass  to the nominal value.
We accept events  with $|m_{\pi^+\pi^0\pi^-\pi^0}-E_{cm}|<0.28$ GeV and 
$|\Delta p|<0.2$ GeV/c, where $E_{cm}$ is the total c.m. energy,
and $\Delta p$ is the momentum difference
between the $\pi^+\pi^0\pi^-\pi^0$ system and the $e^+e^-$ system.

We define the $\rho^{\pm}$ mass intervals as [0.5,1.1] GeV/c$^2$,  
and extract $308\pm 25$   $\rho^+\rho^-$ signal events   in the defined mass region.
The significance is estimated to be 9.5 sigma. 

Assuming $\rho^+\rho^-$ is  produced through  
one photon or $\Upsilon(4S)$, there are three independent helicity 
amplitudes ($\mathrm{F}_{\mathrm{\mu}\mathrm{\nu}}$, $\mathrm{\mu}$/$\mathrm{\nu}$ is the helicity 
of $\rho^+/\rho^-$), 
$\mathrm{F}_{00}$, $\mathrm{F}_{10}$, and 
$\mathrm{F}_{11}$ ($\mathrm{F}_{10}=\mathrm{F}_{-10}=\mathrm{F}_{0\pm1},~\mathrm{F}_{11}=\mathrm{F}_{-1-1}$)~\cite{chung}.
The one-dimensional projections for the decay angles involved  can be expressed as:
\begin{equation}
\frac{dN}{d \cos\theta^* } \propto (\sin^2\theta^* |F_{00}|^2+2 (1+\cos^2\theta^*)|F_{10}|^2 +2  \sin^2\theta^*|F_{11}|^
2)
\label{phietaprodang}
\end{equation}
\begin{equation}
\frac{dN}{d\cos\theta_{\pm}} \propto (\cos^2\theta_{\pm}|F_{00}|^2 + (1+\cos^2\theta_{\pm})|F_{10}|^2+\sin^2\theta_{\pm}
|F_{11}|^2)
\label{phietaphihelicityang}
\end{equation}
\begin{equation}
\frac{dN}{ d\varphi_{\pm}} \propto (|F_{00}|^2 + (4-\cos 2 \varphi_{\pm} )|F_{10}|^2+ 2 |F_{11}|^2)
\label{phietavarphiang}
\end{equation}
where $\theta^*$ is the $\rho$ production angle, $\theta_{\pm}$ $(\varphi_{\pm})$ is the 
helicity (azimuthal) angle of the pion from $\rho$ decay.
From the two dimensional mass fit ($\pi^{+}\pi^0$ and $\pi^{-}\pi^0$), we can 
calculate a $\rho^+\rho^-$ signal sWeight~\cite{sweight}
for each event (including those events outside the defined $\rho^{\pm}$ mass window) 
and  use it to produce signal angular distributions. 
We fit the five angular distributions  to  Equs.~\ref{phietaprodang}, 
~\ref{phietaphihelicityang} and ~\ref{phietavarphiang} 
simultaneously by minimizing $\chi^2$.
The correlations among the five angles are neglected; this is justified by means of fits to events generated according 
the assumed PDFs (toy MC). 
We normalize the amplitudes such that $|F_{00}|^2+4|F_{10}|^2+2|F_{11}|^2=1$ since we have 
1 $F_{00}$, 4 $F_{10}$ and 2 $F_{11}$ amplitude contributions. 
The normalized amplitudes  from the fit are found to be in the ratio:
$\Myamplitudes$, and  $|\mathrm{F_{00}}|^2$  deviates from 1 with  significance more than 3 sigma. This   
disagrees with a QCD prediction~\cite{stanrule}, and suggests that either the decay is not dominated by  single-photon 
exchange as  naively expected, or that the QCD prediction does not apply to data in our energy region. 
The final radiation-corrected cross section for
$0.5~<~m_{\rho^{\pm}}~<~1.1 $ GeV/c$^2$, and 
within $|\cos\theta^*|<$ 0.8, $|\cos\theta_{\pm}|\! <\! 0.85$, 
at near $\sqrt{s}=10.58$ GeV (assuming only one-photon production) is:
\begin{eqnarray*}
 \sigma_{\mathrm{fid}}(e^+e^-\rightarrow\rho^+ \rho^-) & = & \Resultrhoprhom .
\end{eqnarray*}
We extend the cross section calculation from our acceptance region to the full phase space using 
the fitted amplitude values, and find   
$\ResultrhoprhomExtend$; the third uncertainty is due to the amplitude uncertainties.

\section{Conclusion}

We report the first observation of $e^+e^-$ annihilations into hadronic 
states of positive $C$-parity, $\rho^0 \rho^0$  and $\phi\rho^0$. 
We also report the observation of the process $e^+e^-\rightarrow \phi\eta$, and obtain  preliminary results on  
$e^+e^-\rightarrow \rho^+\rho^-$. The measured helicity amplitude magnitudes 
from $e^+e^-\rightarrow \rho^+\rho^-$ contradict a QCD prediction 
at a significance of more than 3 sigma.

\section*{Acknowledgments}
The author wishes to thank his \babar~ colleagues for their support.

\end{document}